\documentclass[sigconf]{acmart}

\renewcommand\footnotetextcopyrightpermission[1]{}
\setcopyright{acmcopyright}
\copyrightyear{2023}
\acmYear{2023}
\acmDOI{XXXXXXX.XXXXXXX}

\acmPrice{15.00}
\acmISBN{978-1-4503-XXXX-X/18/06}

\usepackage{amsmath,amssymb,amsfonts}
\usepackage{algorithm}
\usepackage{algorithmic}
\usepackage{graphicx}
\usepackage{textcomp}
\usepackage{xcolor}
\def\BibTeX{{\rm B\kern-.05em{\sc i\kern-.025em b}\kern-.08em
    T\kern-.1667em\lower.7ex\hbox{E}\kern-.125emX}}

\usepackage{balance}  
\usepackage{multirow}
\usepackage{arydshln}

\usepackage{hyperref}
\usepackage{thumbpdf}
\usepackage{booktabs}
\usepackage{xcolor}
\usepackage{colortbl}
\usepackage{url}

\usepackage{graphicx}
\usepackage{subfigure}
\usepackage{xfrac}
\usepackage{fix-cm}
\usepackage{array}
\usepackage[normalem]{ulem}
\usepackage{xspace}
\usepackage{wrapfig}
\usepackage{tcolorbox}
 
\usepackage[framemethod=tikz]{mdframed}
\usepackage{lipsum}
\usepackage{svg}

\usepackage{enumitem}

\setenumerate[1]{itemsep=0pt,partopsep=0pt,parsep=\parskip,topsep=5pt}
\setitemize[1]{itemsep=0pt,partopsep=0pt,parsep=\parskip,topsep=5pt}
\setdescription{itemsep=0pt,partopsep=0pt,parsep=\parskip,topsep=5pt}

\newcommand\cloud{Microsoft Azure\xspace}
\newcommand\name{\textsc{Uptake}\xspace}
\newcommand\problem{UPLearning\xspace}

\def\ie{\textit{i.e.},~}

\def\etc{\textit{etc.}~}
\def\eg{\textit{e.g.},~}

\def\Snospace~{\S{}}

\definecolor{mygreen3}{rgb}{0, 0.4, 0.05}
\definecolor{haozhegreen}{RGB}{172, 217, 38}

\setcitestyle{numbers,sort&compress}

\setcopyright{acmcopyright}

\newcounter{insightC}

\newcommand{\point}[1]{\vspace{1mm}\noindent\textbf{#1}.}

\newcommand{\fix}[1]{\textcolor{black}{#1}}

\begin{document}

\title{Why does Prediction Accuracy Decrease over Time? \\ Uncertain Positive Learning for Cloud Failure Prediction}

\settopmatter{authorsperrow=1}

\author{Haozhe Li*, Minghua Ma, Yudong Liu, Pu Zhao, Lingling Zheng, Ze Li, Yingnong Dang, \\ Murali Chintalapati, Saravan Rajmohan, Qingwei Lin, Dongmei Zhang}
\thanks{* The work is done during Haozhe Li's internship at Microsoft.}
\affiliation{
  \institution{Microsoft}
  \city{}
  \country{}
}







\renewcommand{\authors}{Haozhe Li, Minghua Ma, Yudong Liu, Pu Zhao, Lingling Zheng, Ze Li, Yingnong Dang, Murali Chintalapati, Saravan Rajmohan, Qingwei Lin, Dongmei Zhang}
\renewcommand{\shortauthors}{H.Li, M.Ma, Y.Liu, P.Zhao, L.Zheng, Z.Li, Y.Dang, M.Chintalapati, S.Rajmohan, Q.Lin, D.Zhang}

\begin{abstract}
With the rapid growth of cloud computing, a variety of software services have been deployed in the cloud. To ensure the reliability of cloud services, prior studies focus on failure instance (disk, node, and switch,  \textit{etc}.) prediction. Once the output of prediction is positive, mitigation actions are taken to rapidly resolve the underlying failure. According to our real-world practice in Microsoft Azure, we find that the prediction accuracy may decrease by about $9\%$ after retraining the models. 
Considering that the mitigation actions may result in uncertain positive instances since they cannot be verified after mitigation, which may introduce more noise while updating the prediction model. 
To the best of our knowledge, we are the first to identify this Uncertain Positive Learning (UPLearning) issue in the real-world cloud failure prediction scenario. To tackle this problem, we design an Uncertain Positive Learning Risk Estimator (Uptake) approach. Using two real-world datasets of disk failure prediction and conducting node prediction experiments in \cloud, which is a top-tier cloud provider that serves millions of users, we demonstrate Uptake can significantly improve the failure prediction accuracy by 5\% on average.

\end{abstract}



\maketitle

\section{Introduction}
\label{sec:intro}
The IT industry has seen a significant trend towards migrating large workloads to the cloud, such as Microsoft Azure, Amazon Web Services, and Google Cloud Platform. 
These providers monitor and analyze thousands of metrics from their cloud infrastructures to ensure high-quality service for billions of users worldwide \cite{babaioff2022truthful, zeng2022fograph, alhilal2022nebula}. 
These metrics can help detect and prevent failures in various components of the cloud system, such as memory \cite{10.1145/3240302.3240309}, disk \cite{LuoNtam21,liu2022multi}, node \cite{lin2018predicting}, or switch \cite{zhang2018prefix}, \textit{etc}.  
Machine learning or deep learning techniques have been applied to these metrics to predict failures and take proactive actions to mitigate them, which can improve the availability and performance of cloud-based software systems while reducing operational costs and risks.

A lot of previous studies aim to design better machine learning models to improve the performance of cloud failure prediction tasks. 
For example, RNN \cite{XuEtAl16}, LSTM \cite{ZhaEtAl18}, Transformer \cite{LuoNtam21}, and TCNN \cite{SunEtAl19} models are used to predict cloud failures. 
When a prediction indicates a problem, we take quick action to fix the underlying issue. 
For example, if a computing node fails, we might move the running virtual machine to another node with minimal disruption or preserve the virtual machine's state during a node reboot \cite{levy2020predictive}. 
Although these actions change the node's status, they do not offer a detailed analysis of the root cause of the failure. 
Consequently, we cannot be certain whether the predicted failure will actually occur, which we call an ``\textit{uncertain positive}'' instance.

According to our experience in deploying cloud failure prediction models in \cloud, we notice that the model updating meets great challenges in real-world online usage.
Model updating is the process of retraining the machine learning model over time (weekly or monthly) to adapt to the changing cloud environment (new hardware and software) \cite{ma2018robust, levy2020predictive}.
An empirical study conducted on both open datasets and \cloud (see \autoref{sec:empirical}), suggests that the prediction accuracy of updated models may decrease by about 9\% over time because the uncertain positive instances introduce noise to the model updating.
This effect is compounded when using a continuously updated model that accumulates uncertain positive instances over time.

\begin{figure}[tp]
    \centering
    \vspace{2mm}
    \includegraphics[width=\linewidth]{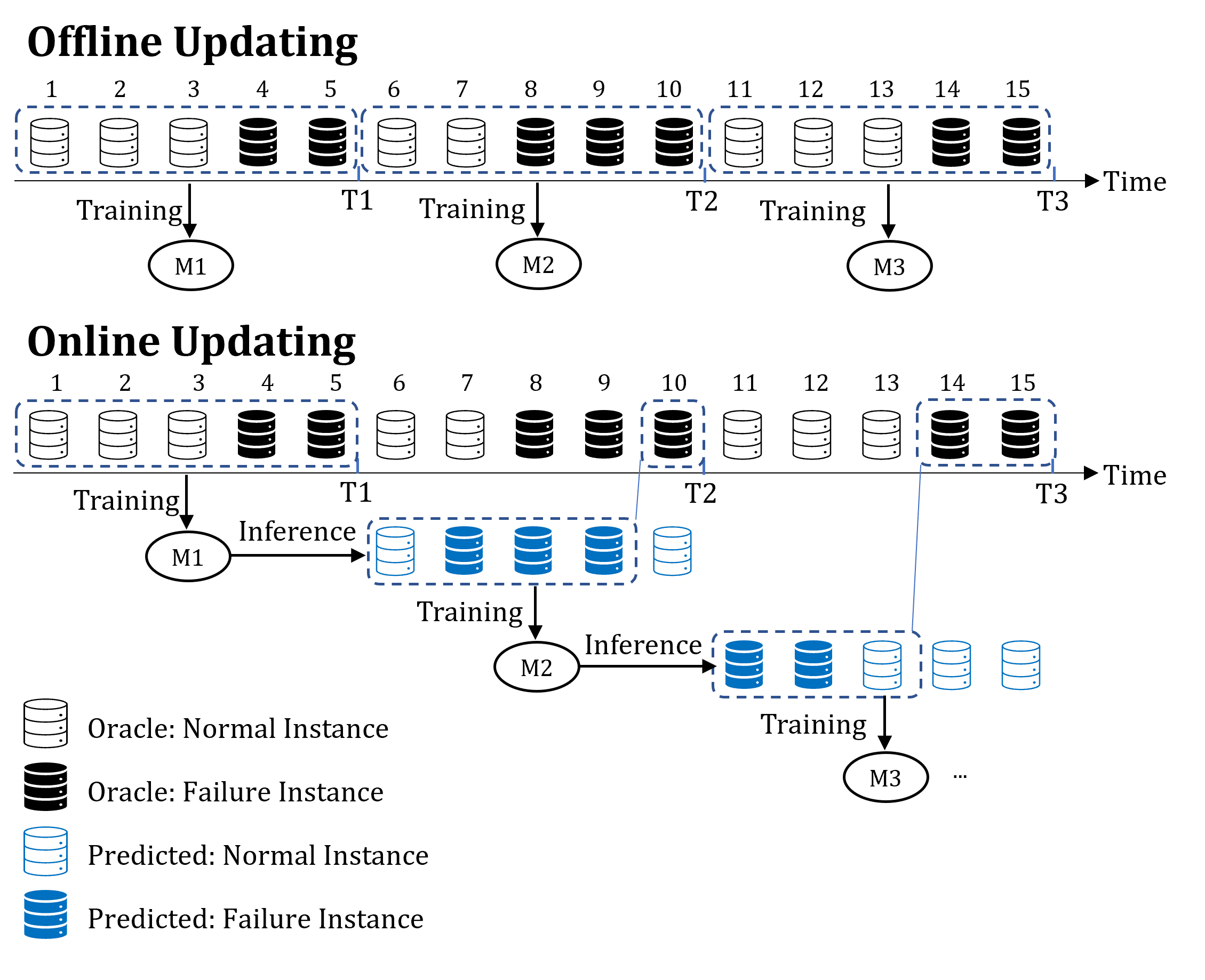}
    \caption{Toy example of cloud failure prediction model (binary classification for normal or failure instances) with offline updating and online updating. The prediction model (M) is updated by training with data in the dashed box in each time stage (T). The number on the top of each sub-figure illustrates the instance ID. }
    \label{fig:motivation}
\end{figure}

More specifically, \autoref{fig:motivation} illustrates a toy example of the machine learning model updating scenario in the cloud failure prediction. 
The figure has two parts, \ie the \textit{offline updating} and the \textit{online updating}. 
The offline updating is an ideal scenario in which we can access the oracle of cloud instance status (shown as the disk icons in black color). 
We may collect the monitoring metrics and status of the cloud instances in each period (T1, T2, T3, \textit{etc.}). 
Then we train the model (M1, M2, M3, \textit{etc.}) in each period without any concern.
When it comes to the online scenario, however, we cannot access the oracle of instances status after T1 (they are identical to the offline scenario). 
In real practice, we train a model in the T1 and make online inferences in the T2 stage, some instances (\#7, \#8, and \#9) are predicted as failure (in blue color). 
Then, these instances are mitigated and cannot access their actual status, which is noted as uncertain positive instances. 
Finally, together with the known failure instance (\#10), we may retrain the model M2. 
Clearly, the \#7 instance is a False Positive one that may introduce noise to model M2. 
Similarly, we may obtain a less accurate model (\#11, \#12 are uncertain positive instances, and \#14, \#15 are true positive instances) in the T3 stage since the model noise may be accumulated.

To the best of our knowledge, we are the first to identify the Uncertain Positive Learning (\problem) challenge in the real-world scenario of cloud failure prediction. 
Addressing this challenge is crucial for ensuring cloud reliability by enabling accurate failure prediction. 
The most closely related research topic to \problem is the  Positive Unlabeled Learning (PULearning) \cite{elkan2008learning,nguyen2011positive,li2009positive,liu2002partially,li2003learning,lee2003learning,liu2003building,du2014analysis,kiryo2017positive}.
PULearning is a machine learning scenario for binary classification where the training set consists of a set of positively labeled instances and an additional unlabeled set that contains positive and negative instances in unknown proportions (so no training instances are explicitly labeled as negative).
The risk estimator-based approach, which estimates the distribution of negative instances in the unlabeled set, is widely used to solve the PULearning problem \cite{kiryo2017positive,li2009positive,nguyen2011positive}.
However, our \problem involves uncertain positive instances, which are similar to the unlabeled set in the PULearning scenario, but also includes both positive and negative instances that system operators investigate after failures occur (see Section \ref{sec:background}). As a result, it is not feasible to directly apply the risk estimator approach used in PULearning.

In this paper, we propose an \textbf{U}ncertain \textbf{P}osi\textbf{t}ive Le\textbf{a}rning Ris\textbf{k} \textbf{E}stimator (\name), a novel approach for various cloud failure prediction models to achieve high prediction accuracy even with uncertain positive instances. 
\name regards uncertain positive instances as both positive and negative through a specially designed risk estimator during the model updating procedure.
Besides, \name only improves the loss function which makes it easy to integrate with various machine learning models, such as RNN, LSTM, Transformer, and TCNN. 
To evaluate the effectiveness of our proposed \name approach, we conduct extensive experiments to compare \name against three model updating approaches on two public disk datasets, Alibaba and Backblaze, which both contain tens of thousands of disks over months. 
We also apply \name to the scenario of node failure prediction in \cloud, which is a top-tier cloud provider and serves millions of customers around the world.
The experiment results on those three datasets demonstrate \name outperforms any other model updating approaches over different prediction models, \ie RNN, LSTM, Transformer, and TCNN.
The F1-scores of \name are $45.26\%$, $70.16\%$, and $69.33\%$ in Alibaba, Backblaze, and \cloud scenarios, which are $4.85\%$, $4.17\%$, and $5.13\%$ better than those achieved by the best baseline approaches, respectively.

To sum up, this work has the following contributions:

\begin{itemize}[leftmargin=*]
    \item To the best of our knowledge, we identify the Uncertain Positive Learning (\problem) problem in the scenario of cloud failure prediction for the first time, which is an important issue to tackle for machine learning model usage in the real-world scenario.
    \item To solve the \problem, we propose an uncertain positive learning risk estimator approach, dubbed \name, which is easy to integrate with various machine learning models. 
    \item To illustrate the generality and effectiveness of \name, we conduct experiments on different cloud failure prediction scenarios, \ie two public disk failure datasets and the node failure from \cloud.
    \name outperforms any other online model updating approach over four widely used failure prediction models. 
    Moreover, \name has been successfully applied to \cloud and improved the reliability of cloud platforms.
\end{itemize}

\section{Background}
\label{sec:background}

\begin{table*}
\centering
\caption{Examples of node \& disk monitoring metrics.}
\label{tab:metrics}
\resizebox{0.85\linewidth}{!}{%
\begin{tabular}{l|l|l} 
\toprule
\textbf{Type} & \textbf{Description} & \textbf{Examples (Feature belongs to Disk/Node)} \\ 
\midrule
Operation Timer & \begin{tabular}[c]{@{}l@{}}The implementation of a timer for various operations of hard disk drives,\\\eg magnetic head seeking and spindle activation, plays a crucial role in \\optimizing the performance of these devices.\end{tabular} & \begin{tabular}[c]{@{}l@{}}\textit{SpinUpTimeNorm }\textbf{(Disk)}\\\textbf{}\textit{SpinRetryCountRaw~}\textbf{(Disk)}\\\textbf{}\textit{Spin\_Retry\_Count\_VALUE} \textbf{(Disk)}\\\textbf{}\textit{Spin\_Up\_Time\_VALUE} \textbf{(Disk)}\end{tabular} \\ 
\midrule
Operation Counter & \begin{tabular}[c]{@{}l@{}}Count of common operations, \eg CPU utilization and power\\consumption, \etc\end{tabular} & \begin{tabular}[c]{@{}l@{}}\textit{PowerCycle~}\textbf{(Node)}\\\textbf{}\textit{PowerOnHours~}\textbf{(Node)}\\\textbf{}\textit{PowerConsumption~}\textbf{(Node)}\end{tabular} \\ 
\midrule
Physical Characteristics & \begin{tabular}[c]{@{}l@{}}The measurement of physical characteristics,~ \eg temperature and\\humidity, \etc\end{tabular} & \begin{tabular}[c]{@{}l@{}}\textit{AirflowTempRaw~}\textbf{(Node)}\\\textbf{}\textit{TemperatureRaw} \textbf{(Node)}\\\textbf{}\textit{Temperature} \textbf{(Node)}\end{tabular} \\ 
\midrule
Error Counter & \begin{tabular}[c]{@{}l@{}}A count of specific, identifiable errors is maintained. This count typically \\remains unchanged unless an error occurs, at which point the count is \\updated to reflect the occurrence of the error.\end{tabular}  & \begin{tabular}[c]{@{}l@{}}\textit{Program\_Fail\_Cnt\_Total\_RAW\_VALUE} \textbf{(Disk)}\textit{}\\\textit{Seek\_Error\_Rate\_RAW\_VALUE~}\textbf{(Disk)}\\\textit{TotalErrors} \textbf{(Node)}\textit{}\\\textit{IOEDCErrorCountRaw~}\textbf{(Node)}\\\textit{MediaErrors~}\textbf{(Node)}\\\textit{ErrorInfoLogEntryCount~}\textbf{(Node)}\end{tabular} \\ 
\midrule
EVENT & \begin{tabular}[c]{@{}l@{}}Windows event are typically used for indicating exception handling for\\Windows servers.~\end{tabular} & \begin{tabular}[c]{@{}l@{}}\textit{WindosStorportEvent\_534 }\textbf{(Node)}\\\textit{WindosStorportEvent\_554~}\textbf{(Node)}\end{tabular} \\
\bottomrule
\end{tabular}
}
\end{table*}

\point{Cloud system} Cloud systems, such as Amazon Web Services (AWS), Microsoft Azure, and Alibaba Cloud, provide pay-as-you-go services for companies and developers, making it convenient for them to deploy their web applications through the Internet. 
These cloud systems serve millions of users around the world, and any failure in the cloud will have a significant impact.
Therefore, it is of great importance to keep the reliability of cloud systems.
Among the reliability efforts in cloud systems, failure prediction, which can detect failure before it actually happens, is one of the most effective solutions and is well-studied in the literature. 

\point{Cloud failure prediction} Cloud failure refers to a state where cloud instances become unavailable due to hardware interruptions, code bugs, or high workload demands. 
Cloud failure can be categorized based on the affected instances, such as node failure \cite{lin2018predicting}, network issue \cite{CheEtAl19}, disk failure \cite{LuoNtam21}, service overloading \cite{10.1145/3267809.3267823}, \etc 
In the following, we introduce two typical cloud failure prediction scenarios, \ie disk failure prediction and node failure prediction.  

\begin{itemize}[leftmargin=*]
    \item Node failures may occur for a variety of reasons (\eg OS crashes, application bugs, misconfigurations, memory leaks, software incompatibility, overheating, and service exceptions). 
    The node failure can be predicted by a wide range of monitoring metrics (signals indicating their temporal and spatial information). 
    MING \cite{lin2018predicting} predicts node failure by taking into account both spatial and temporal signals.
    \autoref{tab:metrics} presents several monitoring metrics of a node, which are classified into six distinct categories: operation timer, physical characteristics, operation counter, error counter, logical input/output, and EVENT data. 
    A detailed explanation of these categories, including examples, is illustrated in \autoref{tab:metrics}. 
    The collection of these metrics is conducted at regular intervals, except for EVENT data. To analyze these metrics, a sliding window technique is employed in the prediction model, resulting in the transformation of the metrics into a feature vector.

    \item Disk failure in cloud systems is a type of hardware failure that can lead to service downtime.
    Disk failure can be predicted based on its status data, known as the SMART (Self-Monitoring, Analysis and Reporting Technology) \cite{enwiki:smart,lu2020making}.
    The monitoring metrics of a disk are SMART (Self-Monitoring, Analysis, and Reporting Technology) \cite{enwiki:smart, lu2020making}. These attributes, which include metrics such as temperature, spin-up time, and reallocated sectors (examples shown in \autoref{tab:metrics}), are employed to predict potential failures. 
    By providing early warning of such failures, SMART technology enables proactive maintenance and replacement to be performed prior to a failure occurring.
\end{itemize}

Based on previous studies, we introduce and implement four widely used machine learning models for cloud failure prediction.

\begin{itemize}[leftmargin=*]
    \item \textbf{RNN} \cite{XuEtAl16}: Recurrent Neural Network is a widely-used deep learning model designed for sequential data. During prediction, RNN uses its recurrent unit to feature the difference between the normal state and failure state from the input sequential data.
    \item \textbf{LSTM} \cite{ZhaEtAl18}: Long Short-Term Memory is an advanced version of recurrent neural network architecture that was designed to model chronological sequences and their long-range dependencies more precisely than conventional RNNs. With long-term features, LSTM can always perform better than RNN and can typically class difficult examples in RNN.
    \item \textbf{Transformer} \cite{LuoNtam21}: The Transformer Model is a novel deep learning-based approach to failure prediction tasks with the attention mechanism, which can capture the temporal nature from instance status data. 
    Transformer utilizes not only a single instance's status data but also considers its neighbors' status data to optimize its prediction performance.
    \item \textbf{TCNN} \cite{SunEtAl19}: Temporal Convolutional Network is a variation of Convolutional Neural Networks for sequence modeling tasks, by combining aspects of RNN and CNN architectures. 
\end{itemize}

\point{Model updating} 
The distribution of online monitoring metrics changes with dynamic software and hardware updates in cloud systems, which causes the distribution learned by the previous model to deviate significantly from the online distribution \cite{ma2018robust}. 
As a result, the prediction performance of the model may degrade.  
To ensure prediction performance, machine learning models deployed online need to be updated over time (weekly or monthly). 
Model updating needs a considerable number of failure instances. 
Updating cloud failure prediction models using pre-training and fine-tuning \cite{Xu_2021_ICCV} strategies [32] is not feasible due to the imbalance between positive and negative instances, as well as the insufficient number of positive instances in real-world scenarios.

\section{An Empirical Study of Prediction Accuracy over Time}
\label{sec:empirical}
We conduct an empirical study on failure prediction.
In this study, we aim to uncover the problem based on our experience in deploying failure prediction models in \cloud. 
We address the following research questions:

\begin{itemize}[leftmargin=*]
    \item RQ1: How does the failure prediction accuracy change over time?
    \item RQ2: Why does the prediction accuracy decrease over time?
\end{itemize}

\subsection{Subjects}

\label{sec:dataset}
We adopt two public datasets \cite{LuoNtam21,liu2022multi} for disk-level failure prediction, \ie Alibaba Cloud and Backblaze datasets.
Besides, \cloud provides large-scale datasets for node-level failure prediction.

\textbf{Alibaba} is collected from large-scale data centers and published by Alibaba Cloud for PAKDD 2020 Alibaba AIOps Competition \cite{ali_tianchi}, which contains millions of disks with a period of more than 16 months. In our experiment, we adopt the dataset within 10 months and split it into five continuous time phases of equal length.
Each time phase contains more than 8,000 disk records with a period of two months and each record has 30 days long monitoring metrics. 
We use 17-dimension features for failure prediction.

\textbf{Backblaze} is a public dataset published by Backblaze, based on the hard drives in Backblaze data center \cite{backblaze2019}. Each disk of the dataset has a label indicating its status. 
In our experiment, we use five months from 2021Q4 to 2022Q1, and it contains over 90,000 disks in total.
We split the dataset into five continuous time phases of equal length. 
Each time phase contains approximately 16,000 disk records and each record contains 38 dimension features with a period of 30 days.

\textbf{\cloud} is a large-scale commercial cloud system that includes node-level monitoring metrics. Over 500,000 node recordings overall from a period of 35 days are employed in our experiment. 
Additionally, we divide all records into five continuous time phases of equal length. 
Each time phase comprises more than 100,000 node records.
The feature input is the 23-dimension monitoring metrics of nodes in a 48-hour time window. 

To demonstrate the generality and robustness of our investigation, we employed different periods for model updating: two months for Alibaba, one month for Backblaze, and one week for \cloud. These periods were selected based on the number of positive instances available in each, as positive instances are significantly less frequent than negative ones in real-world datasets, as noted by Ntam\cite{LuoNtam21}. 
We aim to ensure that each period contains a relatively sufficient number of positive instances, maintaining an imbalance rate (\#Positive to \#Negative ratio) of approximately 1\%. 

\subsection{RQ1: Prediction Accuracy over Time}

\begin{figure}
    \centering
    \includegraphics[width=.85\linewidth]{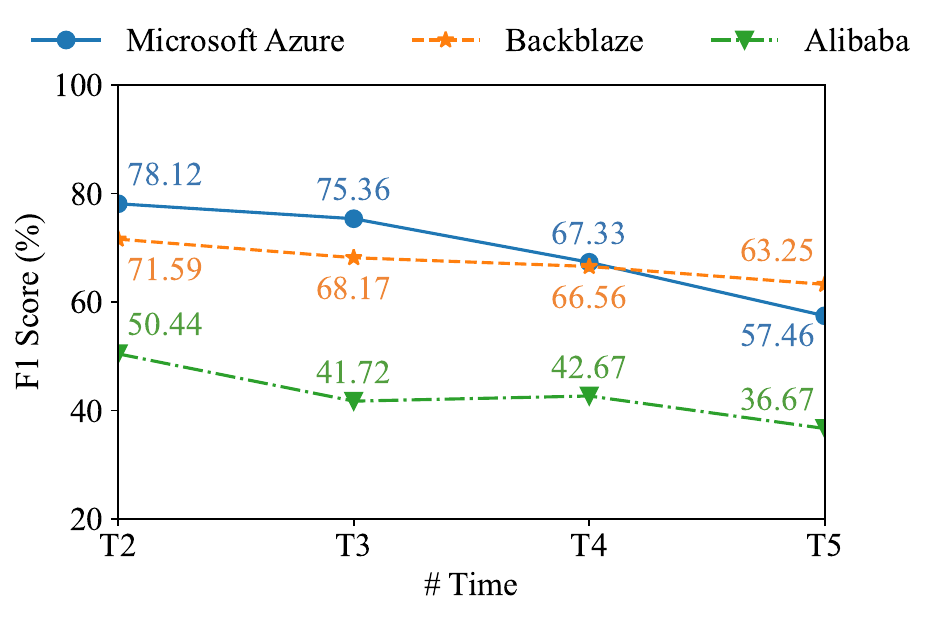}
    \caption{Cloud failure prediction average F1-score over time. }
    \label{fig:empirical}
\end{figure}

\fix{
We undertake an investigation on prediction performance trends across a continuous long range of time, which is divided into 5 equal-length time phases.
At the start of each phase (except for the first \textit{training} phase T1), model updating is performed using monitoring data collected from the previous phase, then deployed for online prediction.
The average prediction performance of the model is calculated at each time phase.
}

\fix{
The prediction F1-score (see \autoref{sec:experiment-evaluation}) on 3 datasets across 4-time phases is illustrated in \autoref{fig:empirical}. 
The time in the figure starts from T2, which is because monitoring data in the T1 phase is used for the initial model training. 
Overall, there is a significant and consistent decrease in the accuracy of the failure prediction model over time for all datasets. Performance on \cloud decreases from 78.22\% on T2 to 57.46\% on T5, a decrease of 20.66\%. Backblaze and Alibaba also show decreases of 8.34\% and 13.77\%, respectively.
The results on Alibaba do not show a decreasing trend in performance on T3 and T4 because the percentage and performance of failures are not completely consistent across time, but the performance of failure prediction on Alibaba also shows a significant decreasing trend of 13.77\% when observed over the entire experimental time.
}

\fix{
The decrease in prediction accuracy over time can be attributed to changes in the distribution of online data. However, it is important to note that this accuracy decrease is observed consistently across all three cloud systems and time phases. Furthermore, the retrained model is trained using the latest collected data, which has the least bias compared to the online data. We refer to this decrease in prediction accuracy after retraining as the \problem (\textbf{U}ncertain \textbf{P}ositive \textbf{Learning}) problem. 
}

\subsection{RQ2: Explanation of Accuracy Decrease}

\label{sec:rq2}

\begin{table}
\centering
\caption{Mitigation actions to cloud failures.}
\label{tab:mitigation}
\resizebox{0.85\linewidth}{!}{%
\begin{tabular}{>{\hspace{0pt}}m{0.25\linewidth}|>{\hspace{0pt}}m{0.69\linewidth}} 
\toprule
Action & Description \\ 
\midrule
Live Migration & Move running VMs from to other failure-free nodes \\ 
\midrule
VM Preserving \par{} Soft Reboot & Reboot the host OS kernel and preserve the VM states \\ 
\midrule
Service Healing & Disconnect current VM and generate new assignment of the VM to healthy nodes \\ 
\midrule
Mark Unallocatable & Block allocation of new VMs to a node \\ 
\midrule
Avoid & Reduce the weight to allocate a new VM to a node \\
\bottomrule
\end{tabular}
}
\end{table}

\fix{
It is mitigation actions that contaminate the training data and lead to a decrease in prediction accuracy (\ie \problem) during model retraining.
In the context of failure prediction in cloud systems, when a cloud component is anticipated to fail, mitigation actions\cite{levy2020predictive} in \autoref{tab:mitigation} are promptly undertaken to either resolve the underlying failure or minimize the impact on services. The effectiveness of these mitigation actions, however, makes it impractical to accurately verify the accuracy of the predictions. 
For instance, in the case of applying live migration to a predicted failing node, all run-time state is migrated to a new node, resulting in the release of workload on the original node\cite{clark2005live}. This operation alters the original state, rendering it unverifiable whether the failure would have indeed occurred in the original state.
}

\begin{figure}
    \centering
    \includegraphics[width=.9\linewidth]{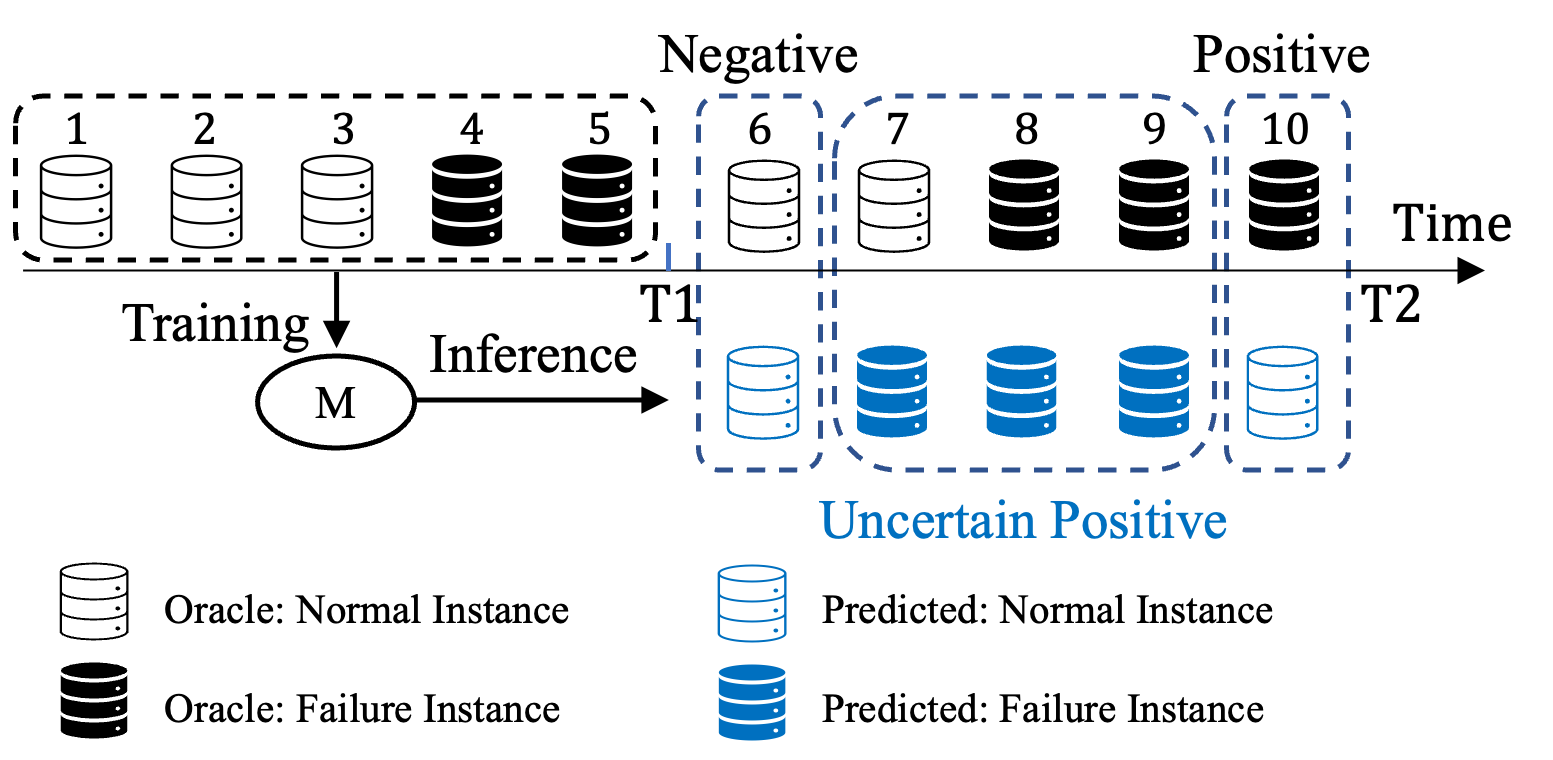}
    \caption{Example of positive, negative, and uncertain positive in the scenario of online cloud failure prediction.}
    \label{fig:explain-up}
\end{figure}

\point{UPLearning problem} 
As we discussed before, the Uncertain Positive Learning (\problem) problem in cloud systems is mainly due to mitigation actions, which change the state of the instances and make it impossible to verify the accuracy of the predictions. 
In the context of cloud failure prediction, an online model (M) is initially trained using historical data with true labels (oracle labels) during period T1. This model predicts instances that are likely to fail, represented by blue-filled instances in period T2, applying mitigation actions to address the potential failure. 
However, it remains unknown whether these predicted instances will actually fail or not, as indicated by the black-bordered instances (\#6-\#10). \textit{Uncertain Positive} instances are those predicted as failures by the online model, such as instances \#7, \#8, and \#9 in \autoref{fig:explain-up}. 
However, the true labels (oracle labels) for these instances (\#7: Negative, \#8 and \#9: Positive) cannot be verified after applying failure mitigation actions, because mitigation actions change the original state of the instances.

\section{Approach}
\label{sec:approach}

We propose a novel model updating approach to solve the \problem problem, named \textbf{U}ncertain \textbf{P}osi\textbf{t}ive Le\textbf{a}rning Ris\textbf{k} \textbf{E}stimator (\name), which still consumes uncertain positive instances during model updating and significantly improves the online model updating performance for cloud failure prediction.

In this section, 

we first present the problem formulation of \problem in \autoref{sec:approach-problem}. Then we introduce the technical details of \name in \autoref{sec:approach-up-risk}.

\subsection{Problem Settings}
\label{sec:approach-problem}

Before diving into the approach details, we present a formal definition of the cloud failure prediction task.
We define the cloud failure prediction task as a binary classification problem based on monitoring metrics. 
Essentially, we collect feature vectors with $d$ dimensions, which contain different types of monitoring metrics from cloud infrastructures, at regular intervals (e.g. hourly or daily). These feature vectors are used to create a continuous time series of features, denoted as $X = [x_1, x_2, ..., x_l] \in \mathbb{R}^d$, where $l$ is the number of timestamps in the feature $X$. 
For each feature $X$, we assign a label $Y \in {0, 1}$ based on whether it represents a failure instance $(Y = 1)$ or a normal instance $(Y = 0)$ in the cloud system. The goal of cloud failure prediction is to train a classifier $f_X \to Y$ that can predict the class label $Y$ when given a feature $X$ with uncertain positive labels.

The training set is a collection of features $X$ with their corresponding label $Y$, denoted as $\mathcal{N}=\{ (X_1, Y_1), ..., (X_n, Y_n) \}$, where $n=\left| \mathcal{N}\right|$. Our target is to minimize prediction loss $\mathcal{L}$, (\eg the binary cross-entropy loss), on the training set.

In the scenario of cloud failure prediction, some labels $Y_i (i\in\{1, ... , n\})$ in the training set are uncertain, specifically, we have to use the training set which contains uncertain label $\hat{Y}$ to train and update the online model. The training loss to be minimized is represented as:
\begin{equation}
    \mathcal{L}= - \frac{1}{n}\sum^{n}_{i=1}l(f(X_i), \hat{Y_i})
    \label{equ:uncertain-loss}
\end{equation}
where $l$ is the loss function, $f$ is the binary classifier, and $(X_i, \hat{Y}_i),\quad i \in \{1, ... n\}$ is the sample from the training set.
$\hat{Y}$ is the uncertain label (\ie the predicted label from the online model).

\subsection{\name}
\label{sec:approach-up-risk}

Typically, the model's performance is assessed using the risk function \cite{NIPS2014_35051070}. A precise risk function enhances the performance of the failure prediction model. In this section, we introduce a new method for cloud failure prediction called \name (Uncertain Positive Learning Risk Estimator). This approach ensures high prediction accuracy, especially when handling uncertain positive instances, and effectively addresses the \problem problem.

\point{Risk estimator}
In this section, we describe the binary classification representation of the risk function \cite{NIPS2014_35051070,pmlr-v37-plessis15}. Consider a feature set \(X\in \mathbb{R}^d\) and its corresponding label \(Y\in\{0, 1\}\). Let \(p(x,y)\) be the underlying joint density of \((X, Y)\), \(p_{\rm p}(x) = p(x|Y=1)\), and \(p_{\rm n}(x)=p(x|Y=0)\) represent the positive and negative marginals, respectively. Additionally, define \(\pi_{\rm p} = p(Y = 1)\) as the class-prior probability and \(\pi_{\rm n} = p(Y = 0) = 1-\pi_{\rm p}\).

Let \(f_{X}\to Y\) be the decision function, and \(l: \mathbb{R} \times \{0, 1\} \to \mathbb{R}\) be the loss function, where \(l(f(X),Y)\) measures the loss between the predicted output \(f(X)\) and the actual label \(Y\).

Define \(R(f)\) as the risk function for the given decision function \(f_{X}\to Y\), \(R^+_{\rm p}(f)=\mathbb{E}_{\rm p}[l(f(X), 1)]\), and \(R^-_{\rm n}(f)=\mathbb{E}_{\rm n}[l(f(X), 0)]\), where \(\mathbb{E}_{\rm p}[\cdot] = \mathbb{E}_{X \sim p_{\rm p}}[\cdot]\) and \(\mathbb{E}_{\rm n}[\cdot] = \mathbb{E}_{X \sim p_{\rm n}}[\cdot]\), denoting the expectations of positive and negative variables.

In standard binary classification, the risk of the decision function \(f_X\to Y\) is denoted as:

\begin{equation}
    R(f) = \mathbb{E}_{(X, Y)\sim p(x, y)}[l(f(X), Y)] = \pi_{\rm p} R^+_{\rm p}(f)+ \pi_{\rm n} R^-_{\rm n}(f)
    \label{equ:risk-estimate}
\end{equation}

\point{Uncertain Positive Learning risk estimator}
In the context of cloud failure prediction, we encounter instances not only labeled as positive or negative but also instances with uncertain positive labels predicted by the online model. \name treats uncertain positive instances as having both positive and negative labels with different weights. The risk function for this scenario is represented as follows:

\begin{equation}
    R(f) = \pi_{\rm p} R^+_{\rm p}(f) + \pi_{\rm n} R^-_{\rm n}(f) + (\pi_{\rm p} R^+_{\rm u}(f) + \pi_{\rm n}  R^-_{\rm u}(f))
\end{equation}

Here, \(R_{\rm u}^+(f) =\mathbb{E}_{\rm u}[l(f(X), 1)]\) and \(R_{\rm u}^-(f) =\mathbb{E}_{\rm u}[l(f(X), 0)]\) represent the expectations of the decision function \(f(X)\) when dealing with uncertain positive instances.

It's crucial to note that \(\pi_{\rm p}\) and \(\pi_{\rm n}\) are hyper-parameters in this study. \(\pi_{\rm p}\) estimates the proportion of real positive instances among uncertain positive instances, while \(\pi_{\rm n} = 1 - \pi_{\rm p}\) estimates the percentage of negative instances. In this paper, \(\pi_{\rm p}\) is set to the Precision value obtained during the T1 validation stage of hyper-parameter tuning (refer to Section \ref{sec:experiment-parameter}). When \(\pi_{\rm p}\) approximates the proportion of uncertain positive instances with true positive labels, \name demonstrates improved performance. The Precision value from the T1 validation stage offers an accurate estimate of the proportion of uncertain positive instances with positive labels.

\point{Implementation in various models} 
\name is a generic approach that can be integrated with any machine learning model for cloud failure prediction.

Algorithm \ref{alg:uplearning} illustrates the core implementation of \name. The input training data subscripted with $\rm p$, $\rm n$, and $\rm u$ are the instances labeled with positive, negative, and uncertain positive, respectively, and $\theta$ indicates the model parameter for decision function $f(X;\theta)$. 
To reduce the interference of uncertain positive instances to the direction of gradient descent, \name improves the loss function in the model updating procedure by adjusting the direction of the gradient. 

Since \name only improves the original loss function without changing its property, it is suitable for various machine learning models.

\begin{algorithm}
    \caption{Introducing \name in the gradient descent}
    \label{alg:uplearning}
    \renewcommand{\algorithmicrequire}{\textbf{Input:}}
    \renewcommand{\algorithmicensure}{\textbf{Output:}}
    \begin{algorithmic}[1]
        \REQUIRE training data ($X_{\rm p}, X_{\rm n}, X_{\rm u}$); \\ hyperparameter $0\leq \pi_{\rm p} \leq 1$, and $\pi_{\rm n} = 1 - \pi_{\rm p}$;\\ decision function $f(X;\theta)$; loss function $l(f(X),Y)$
        \ENSURE gradient $\nabla_{\theta}$ to update $\theta$ for $f(X;\theta)$
        \STATE \textit{Improved loss}  $\mathcal{L} = \pi_{\rm p} l(f([X_{\rm p}, X_{\rm u}]), 1) + \pi_{\rm n} l(f([X_{\rm n}, X_{\rm u}]), 0)$
        \STATE Set gradient $\nabla_\theta \mathcal{L}$ and update $\theta$
    \end{algorithmic}
\end{algorithm}

\section{Experiment}
\label{sec:experiment}
In this section, we conduct extensive experiments to demonstrate the effectiveness of our \name approach when integrated with existing failure prediction models. 
First, we present the experimental settings, including datasets, the failure prediction models, and three compared approaches. 
\fix{Then, we conduct experiments to verify the following research questions:
\begin{itemize}[leftmargin=*]
    \item \textbf{RQ3:} How does \name perform for \problem problem in cloud failure prediction?
    \item \textbf{RQ4:} How does \name perform in the scenario of online cloud systems?
    \item \textbf{RQ5:} Does \name exhibit consistent performance across different base models?
    \item \textbf{RQ6:} What is the impact of \name on predicting efficiency?
    \item \textbf{RQ7:} How \name's parameters impact its prediction performance?
\end{itemize}
RQ3 quantitatively measures the performance of \name and compares it with several in the public disk prediction datasets. RQ4 applies \name to a large-scale commercial cloud system to evaluate its performance in a real-world setting.
}

\subsection{Settings}

\subsubsection{Datasets}

To evaluate the reliability and applicability of \name, we employ two publicly accessible datasets: the Alibaba Cloud and Backblaze datasets \cite{LuoNtam21, liu2022multi}, for predicting disk failures. Furthermore, we integrate a vast industrial cloud system dataset to predict node failures. These datasets together form the foundation of our assessment and analysis. The specific configuration and details of these datasets align with what we outlined in Section \ref{sec:dataset}.

\subsubsection{Implementations and Environments}
\label{sec:implement}
We implement all four deep learning models, \ie RNN, LSTM, Transformer, and TCNN, based on Python 3.8.13 and PyTorch 1.11.0 \cite{NEURIPS2019_9015}. We use identical parameter values as in the previous works for the compared approaches. The model was selected based on the epoch that achieved the highest F1-score on the validation set.

We conduct all the experiments on Linux Dev Node with Ubuntu 20.04 LTS 64-bit operating system, 24 cores AMD Epyc 7V13, 220 GB memory, and NVIDIA Tesla A100 with 80 GB GPU memory.

\subsubsection{Important Hyper-parameters}
\label{sec:parameters}
As explained in Section \ref{sec:approach-up-risk}, the parameter \(\pi_{\rm p}\) plays a crucial role in this study. Its value is derived from the Precision metric obtained during the T1 validation stage.

\subsubsection{Evaluation metrics}
\label{sec:experiment-evaluation}
The cloud failure prediction task is a binary classification task, aligning with existing research. The performance of \name is assessed using Precision, Recall, and F1-score, following standard conventions. Instances are labeled as either normal or failure based on their practical performance, with failure considered positive and normal considered negative. True Positive (TP) represents the number of correctly predicted failure instances, while True Negative (TN) represents the accurately predicted normal instances. Conversely, False Positive (FP) and False Negative (FN) indicate instances that were incorrectly predicted as normal and failure, respectively.

\subsubsection{Compared Approaches}
\label{sec:experiment-compared}
In the experiment section, we compare \name with two model updating approaches, which are named offline updating (offline), 

and online updating with certain positive labels (certain). To the best of our knowledge, \problem has not been proposed before.
Therefore, there are no state-of-the-art solutions for model updating. 
We design the two model updating approaches below as compared with \name.

\begin{itemize}[leftmargin=*]
\item \textbf{\textit{Offline} updating} employs ground truth labels for model training, which contain no uncertain labels. However, this approach is not feasible in practice because part of the ground truth labels are uncertain positive labels. 
It serves as an \textit{upper bound} of the performance for \name and other subsequent approaches.

\item \textbf{Online updating with \textit{certain} positive labels} accumulates all records with certain labels for model updating. For instance, when updating the model for predicting phase T4, the \textit{certain} approach collects all records with certain labels from phase T1 to phase T3. This includes all records in phase T1, as well as true negative (TN) and false negative (FN) records in phases T2 and T3.

\end{itemize}

\begin{table*}[ht]
    \centering
\caption{Performance comparison of \name and its three compared approaches on Alibaba dataset. P, R, and F1 denote precision, recall and F1-score, respectively. }
\label{tab:result-ali}
\resizebox{0.9\linewidth}{!}{%
\begin{tabular}{c|l|ccc|ccc|ccc|ccc|c} 
\toprule
\multirow{2}{*}{Model} & \multirow{2}{*}{Approach} & \multicolumn{3}{c|}{T2} & \multicolumn{3}{c|}{T3} & \multicolumn{3}{c|}{T4} & \multicolumn{3}{c|}{T5} & Avg. \\ 
\cmidrule(lr){3-15}
 &  & P & R & F1 & P & R & F1 & P & R & F1 & P & R & F1 & F1 \\ 
\midrule
\multirow{3}{*}{RNN} & Offline & \multirow{3}{*}{57.84} & \multirow{3}{*}{33.33} & \multirow{3}{*}{42.29} & 40.12 & 50.00 & 44.52 & 76.92 & 35.71 & 48.78 & 43.89 & 49.07 & 46.33 & 46.54 \\ 
\cdashline{2-2}\cdashline{6-15}
 & Certain &  &  &  & 35.95 & 42.31 & 38.87 & 79.59 & 23.21 & 35.94 & 49.12 & 17.39 & 25.69 & 33.50 \\
 & \name &  &  &  & 40.30 & 41.54 & \textbf{40.91} & 47.65 & 42.26 & \textbf{44.79} & 35.96 & 39.75 & \textbf{37.76} & \textbf{41.15} \\ 
\midrule
\multirow{3}{*}{LSTM} & Offline & \multirow{3}{*}{65.31} & \multirow{3}{*}{36.16} & \multirow{3}{*}{46.55} & 65.82 & 40.00 & 49.76 & 74.71 & 38.69 & 50.98 & 63.11 & 40.37 & 49.24 & 49.99 \\ 
\cdashline{2-2}\cdashline{6-15}
 & Certain &  &  &  & 54.22 & 34.62 & 42.25 & 80.43 & 22.02 & 34.58 & 59.09 & 24.22 & 34.36 & 37.06 \\
 & \name &  &  &  & 56.18 & 38.46 & 45.66 & 64.76 & 40.48 & \textbf{49.82} & 58.72 & 39.75 & \textbf{47.41} & \textbf{47.63} \\ 
\midrule
\multirow{3}{*}{\begin{tabular}[c]{@{}c@{}}Trans\\ -former\end{tabular}} & Offline & \multirow{3}{*}{62.41} & \multirow{3}{*}{46.89} & \multirow{3}{*}{53.55} & 61.80 & 42.31 & 50.23 & 68.47 & 45.24 & 54.48 & 61.42 & 48.45 & 54.17 & 52.96 \\ 
\cdashline{2-2}\cdashline{6-15}
 & Certain &  &  &  & 53.92 & 42.31 & 47.41 & 79.66 & 27.98 & 41.41 & 61.76 & 26.09 & 36.68 & 41.83 \\
 & \name &  &  &  & 53.04 & 46.92 & \textbf{49.80} & 46.70 & 50.60 & \textbf{48.57} & 41.24 & 49.69 & \textbf{45.07} & \textbf{47.81} \\ 
\midrule
\multirow{3}{*}{TCNN} & Offline & \multirow{3}{*}{57.72} & \multirow{3}{*}{40.11} & \multirow{3}{*}{47.33} & 43.59 & 52.31 & 47.55 & 60.38 & 38.10 & 46.72 & 58.88 & 39.13 & 47.01 & 47.09 \\ 
\cdashline{2-2}\cdashline{6-15}
 & Certain &  &  &  & 46.25 & 28.46 & 35.24 & 72.41 & 25.00 & 37.17 & 42.86 & 22.36 & 29.39 & 33.93 \\
 & \name &  &  &  & 63.24 & 33.08 & \textbf{43.43} & 36.84 & 54.16 & \textbf{43.85} & 68.29 & 34.78 & \textbf{46.09} & \textbf{44.46} \\ 
\midrule
\multirow{3}{*}{Avg.} & Offline & \multirow{3}{*}{60.82} & \multirow{3}{*}{39.12} & \multirow{3}{*}{47.43} & 52.83 & 46.16 & 48.02 & 70.12 & 39.44 & 50.24 & 56.83 & 44.26 & 49.19 & 49.15 \\ 
\cdashline{2-2}\cdashline{6-15}
 & Certain &  &  &  & 47.59 & 36.93 & 40.94 & 78.02 & 24.55 & 37.28 & 53.21 & 22.52 & 31.53 & 36.58 \\
 & \name &  &  &  & 53.19 & 40.00 & \textbf{44.95} & 48.99 & 46.88 & \textbf{46.76} & 51.05 & 40.99 & \textbf{44.08} & \textbf{45.26} \\
\bottomrule
\end{tabular}
}
\end{table*}

\begin{table*}[ht]
\centering
\caption{Performance comparison of \name and its two compared approaches on Backblaze dataset. P, R, and F1 denote precision, recall and F1-score, respectively.}
\label{tab:result-backblaze}
\resizebox{0.9\linewidth}{!}{%
\begin{tabular}{c|l|ccc|lll|lll|lll|l} 
\toprule
\multirow{2}{*}{Model} & \multicolumn{1}{c|}{\multirow{2}{*}{Approach}} & \multicolumn{3}{c|}{T2} & \multicolumn{3}{c|}{T3} & \multicolumn{3}{c|}{T4} & \multicolumn{3}{c|}{T5} & \multicolumn{1}{c}{Avg.} \\ 
\cmidrule(lr){3-15}
 & \multicolumn{1}{c|}{} & P & R & F1 & \multicolumn{1}{c}{P} & \multicolumn{1}{c}{R} & \multicolumn{1}{c|}{F1} & \multicolumn{1}{c}{P} & \multicolumn{1}{c}{R} & \multicolumn{1}{c|}{F1} & \multicolumn{1}{c}{P} & \multicolumn{1}{c}{R} & \multicolumn{1}{c|}{F1} & \multicolumn{1}{c}{F1} \\ 
\midrule
\multirow{3}{*}{RNN} & Offline & \multirow{3}{*}{50.88} & \multirow{3}{*}{84.67} & \multirow{3}{*}{63.56} & 62.75 & 73.56 & 67.73 & 74.29 & 73.24 & 73.76 & 76.33 & 62.32 & 68.62 & 70.04 \\ 
\cdashline{2-2}\cdashline{6-15}
 & Certain &  &  &  & 57.89 & 50.57 & 53.99 & 80.91 & 41.78 & 55.11 & 83.67 & 39.61 & 53.77 & 54.29 \\ 
\cdashline{15-15}
 & \name &  &  &  & 59.31 & 69.54 & \textbf{64.02} & 62.21 & 87.32 & \textbf{72.66} & 57.73 & 81.16 & \textbf{67.47} & \textbf{68.05} \\ 
\midrule
\multirow{3}{*}{LSTM} & Offline & \multirow{3}{*}{63.59} & \multirow{3}{*}{85.40} & \multirow{3}{*}{72.90} & 70.79 & 82.18 & 76.06 & 76.71 & 78.87 & 77.78 & 70.71 & 81.64 & \textbf{7}5.78 & 76.54 \\ 
\cdashline{2-2}\cdashline{6-15}
 & Certain &  &  &  & 71.81 & 61.49 & 66.25 & 86.79 & 43.19 & 57.68 & 82.86 & 42.03 & 55.77 & 59.90 \\
 & \name &  &  &  & 63.32 & 83.33 & \textbf{71.96} & 61.72 & 74.18 & \textbf{67.38} & 54.79 & 80.19 & \textbf{65.10} & \textbf{68.15} \\ 
\midrule
\multirow{3}{*}{\begin{tabular}[c]{@{}c@{}}Trans\\ -former\end{tabular}} & Offline & \multirow{3}{*}{71.60} & \multirow{3}{*}{84.67} & \multirow{3}{*}{77.59} & 74.87 & 85.63 & 79.89 & 76.23 & 79.81 & 77.98 & 72.44 & 78.74 & 75.46 & 77.78 \\ 
\cdashline{2-2}\cdashline{6-15}
 & Certain &  &  &  & 69.54 & 69.54 & 69.54 & 83.48 & 45.07 & 58.54 & 82.41 & 43.00 & 56.51 & 61.53 \\
 & \name &  &  &  & 71.36 & 87.36 & \textbf{78.55} & 71.31 & 81.69 & \textbf{76.15} & 64.77 & 82.61 & \textbf{72.61} & \textbf{75.77} \\ 
\midrule
\multirow{3}{*}{TCNN} & Offline & \multirow{3}{*}{63.54} & \multirow{3}{*}{83.94} & \multirow{3}{*}{72.33} & 67.35 & 75.86 & 71.35 & 74.24 & 79.81 & 76.92 & 79.63 & 62.32 & 69.92 & 72.73 \\ 
\cdashline{2-2}\cdashline{6-15}
 & Certain &  &  &  & 77.45 & 45.40 & 57.25 & 79.07 & 47.89 & 59.65 & 80.92 & 51.21 & 62.72 & 59.87 \\
 & \name &  &  &  & 59.26 & 73.56 & \textbf{65.64} & 62.45 & 81.22 & \textbf{70.61} & 63.49 & 77.29 & \textbf{69.72} & \textbf{68.66} \\ 
\midrule
\multirow{3}{*}{Avg.} & Offline & \multirow{3}{*}{62.40} & \multirow{3}{*}{84.67} & \multirow{3}{*}{71.60} & 68.94 & 79.31 & 73.76 & 75.37 & 77.93 & 76.61 & 74.78 & 71.26 & 72.45 & 74.27 \\ 
\cdashline{2-2}\cdashline{6-15}
 & Certain &  &  &  & 69.17 & 56.75 & 61.76 & 82.56 & 44.48 & 57.75 & 82.47 & 43.96 & 57.19 & 58.90 \\
 & \name &  &  &  & 63.31 & 78.45 & \textbf{70.04} & 64.42 & 81.10 & \textbf{71.70} & 60.20 & 80.31 & \textbf{68.73} & \textbf{70.16} \\
\bottomrule
\end{tabular}
}
\end{table*}

\subsection{RQ3: Performance of \name}

\label{sec:experiment-rq1}

\begin{table*}
    \centering
\caption{Performance comparison of \name and its two compared approaches on \cloud. P, R, and F1 denote precision, recall and F1-score, respectively. }
\label{tab:result-azure}
\resizebox{0.9\linewidth}{!}{%
\begin{tabular}{c|l|ccc|ccc|ccc|ccc|c} 
\toprule
\multirow{2}{*}{Model} & \multirow{2}{*}{Approach} & \multicolumn{3}{c|}{T2} & \multicolumn{3}{c|}{T3} & \multicolumn{3}{c|}{T4} & \multicolumn{3}{c|}{T5} & Avg. \\ 
\cmidrule(lr){3-15}
 &  & P & R & F1 & P & R & F1 & P & R & F1 & P & R & F1 & F1 \\ 
\midrule
\multirow{3}{*}{RNN} & Offline & \multirow{3}{*}{88.31} & \multirow{3}{*}{86.96} & \multirow{3}{*}{75.16} & 88.59 & 67.11 & 76.36 & 71.43 & 59.14 & 64.71 & 82.24 & 68.75 & 74.89 & 71.99 \\ 
\cdashline{2-2}\cdashline{6-15}
 & Certain &  &  &  & 93.94 & 46.62 & 62.31 & 83.16 & 43.82 & 57.39 & 93.50 & 48.70 & 64.04 & 61.25 \\
 & \name &  &  &  & 94.62 & 49.62 & \textbf{65.10} & 84.65 & 51.88 & \textbf{64.33} & 93.28 & 57.81 & \textbf{71.38} & \textbf{66.94} \\ 
\midrule
\multirow{3}{*}{LSTM} & Offline & \multirow{3}{*}{80.60} & \multirow{3}{*}{79.67} & \multirow{3}{*}{80.13} & 66.19 & 78.76 & 71.93 & 77.44 & 68.28 & 72.57 & 80.79 & 63.54 & 71.14 & 71.88 \\ 
\cdashline{2-2}\cdashline{6-15}
 & Certain &  &  &  & 47.45 & 61.28 & 53.49 & 78.95 & 56.45 & 65.83 & 68.65 & 93.69 & 68.65 & 62.66 \\
 & \name &  &  &  & 59.58 & 79.51 & \textbf{68.12} & 65.05 & 75.54 & \textbf{69.90} & 84.00 & 60.16 & \textbf{70.11} & \textbf{69.38} \\ 
\midrule
\multirow{3}{*}{\begin{tabular}[c]{@{}c@{}}Trans\\ -former\end{tabular}} & Offline & \multirow{3}{*}{65.60} & \multirow{3}{*}{96.55} & \multirow{3}{*}{78.12} & 75.27 & 78.95 & 77.06 & 76.45 & 67.20 & 71.53 & 71.03 & 73.44 & 72.22 & 73.60 \\ 
\cdashline{2-2}\cdashline{6-15}
 & Certain &  &  &  & 80.54 & 56.02 & 66.08 & 82.63 & 52.42 & 64.14 & 91.28 & 51.82 & 66.11 & 65.44 \\
 & \name &  &  &  & 72.38 & 77.82 & \textbf{75.54} & 65.37 & 72.04 & \textbf{68.54} & 71.74 & 68.75 & \textbf{70.21} & \textbf{71.43} \\ 
\midrule
\multirow{3}{*}{TCNN} & Offline & \multirow{3}{*}{80.87} & \multirow{3}{*}{75.70} & \multirow{3}{*}{78.20} & 67.89 & 79.89 & 73.40 & 84.69 & 66.94 & 74.77 & 64.25 & 76.30 & 69.76 & 72.64 \\ 
\cdashline{2-2}\cdashline{6-15}
 & Certain &  &  &  & 45.24 & 57.14 & 50.50 & 86.51 & 50.00 & 63.37 & 91.08 & 50.52 & 64.99 & 59.62 \\
 & \name &  &  &  & 63.72 & 77.26 & \textbf{69.84} & 71.28 & 73.39 & \textbf{72.32} & 57.77 & 78.39 & \textbf{66.52} & \textbf{69.56} \\ 
\midrule
\multirow{3}{*}{Avg.} & Offline & \multirow{3}{*}{78.85} & \multirow{3}{*}{84.72} & \multirow{3}{*}{77.90} & 74.49 & 76.18 & 74.69 & 77.50 & 65.39 & 70.90 & 74.58 & 70.51 & 72.00 & 72.53 \\ 
\cdashline{2-2}\cdashline{6-15}
 & Certain &  &  &  & 66.79 & 55.27 & 58.10 & 82.81 & 50.67 & 62.68 & 86.13 & 61.18 & 65.95 & 62.24 \\
 & \name &  &  &  & 72.58 & 71.05 & \textbf{69.65} & 71.59 & 68.21 & \textbf{68.77} & 76.70 & 66.28 & \textbf{69.56} & \textbf{69.33} \\
\bottomrule
\end{tabular}
}
\end{table*}

We assess our proposed approach, \name, using two public datasets referred to as Alibaba and Backblaze. The results are presented in Table \ref{tab:result-ali} and Table \ref{tab:result-backblaze} respectively. In these tables, the last column displays the average F1-score for each row. The final three rows depict the average results of all four models across different time phases and using different approaches. 

In the rest of the table, every set of three columns showcases the performance metrics (Precision, Recall, and F1-score) for different time phases. Notably, the time phases in Table \ref{tab:result-ali} and Table \ref{tab:result-backblaze} commence from T2, as the data in T1 is utilized for the initial model training. It's important to observe that the three approaches yield identical results at T2 since it marks the first stage to obtain model inference results without encountering any \problem problem (as illustrated in Figure \ref{fig:motivation}). 

The performance of these models remains competitive with previous work \cite{liu2022multi}, considering the inherent challenges of cloud failure prediction tasks in real-world scenarios. It's worth mentioning that online disk failure prediction often faces limitations due to low F1-scores. This challenge largely stems from the prevalence of missing data and significant imbalances in the data distribution, posing significant obstacles to improvement.

Upon analyzing Table \ref{tab:result-ali} and Table \ref{tab:result-backblaze}, we observe that \name performs admirably on these two public datasets. The average F1-scores are $45.26\%$ for Alibaba and $70.16\%$ for Backblaze.
According to the results shared above, we have the following findings.
The results of various methods applied to Alibaba's data demonstrate low performance due to the complexity of the dataset. The highest F1-score achieved on the leader-board is only $49.07\%$. \footnote{https://tianchi.aliyun.com/competition/entrance/231775/rankingList}

\point{Comparison with different updating approaches} 
When examining the last column of Table \ref{tab:result-ali} and Table \ref{tab:result-backblaze}, it becomes evident that \textit{offline} updating consistently outperforms other approaches. It achieves an average F1-score of $49.15\%$ in Alibaba and $74.27\%$ in Backblaze. However, it's important to note that offline updating, while showing the best performance, is not practically applicable as it requires access to all certain labels, which is often unavailable in real-world scenarios. Therefore, we present this approach as the "upper bound" of the prediction model, representing the best possible performance using all certain labels.
Among the practical and feasible approaches, \name demonstrates the best performance. It achieves an average F1-score of $45.26\%$ in Alibaba and $70.16\%$ in Backblaze. In contrast, \textit{Certain} performs the worst, with an average F1-score of $36.58\%$ in Alibaba and $58.90\%$ in Backblaze. We attribute the lower accuracy of \textit{Certain} to the accumulation of phases, which causes the proportion of positive and negative instances to deviate from the actual distribution. Specifically, the \textit{Certain} method disregards all uncertain positive samples, even though only an average of $60.84\%$ in Alibaba and $19.27\%$ in Backblaze of positive labels are certain positive. This discrepancy in handling uncertain positive samples results in a distribution drift, leading to worse performance.

\point{Model comparison}
Analyzing the \name results for each model displayed in Table \ref{tab:result-ali} and Table \ref{tab:result-backblaze}, we observe that \name consistently demonstrates different performance improvements compared to the \textit{Certain} approach across various models. In general, each model in our experiment experiences enhancements when using \name. Specifically, RNN exhibits the most substantial improvement, achieving an average F1-score increase of $7.02\%$ across all datasets, while LSTM shows the smallest improvement, with an average F1-score increase of $1.64\%$. 
However, when compared with offline updating, there isn't a consistent pattern across all datasets concerning model differences. This discrepancy can be attributed to biases introduced in the data distribution among different stages, affecting the performance of various models differently.
In summary, \name consistently outperforms \textit{Certain} model updating approaches and approaches the performance of the \textit{offline} (\ie the upper bound) across different prediction models. This consistency underscores its generality and robustness in enhancing model performance.

\subsection{RQ4: Online Performance}

We have deployed \name to one of the top-tier cloud systems in the world, \cloud, which suffer from the \problem problem before using our approach. 
We have conducted an online experiment (in the A/B testing environment to obtain ground truth) for a period of time over five weeks from July 2022 to August 2022. 

\autoref{tab:result-azure} presents a comparison of model performance for the mentioned model updating approaches, following the same format as the tables in \autoref{sec:experiment-rq1}. \name outperforms \textit{Certain}, achieving an average F1-score of $69.33\%$. This performance surpasses Certain by an average of $7.09\%$ and falls slightly below the \textit{Offline} performance by $3.2\%$ in terms of F1-score.

The accumulation of uncertain positive labels significantly impacts the performance of node failure prediction. Unlike failure prediction on public disk datasets, the node failure prediction model is more concerned with the quality of labels rather than the proportion of positive samples to negative samples. 
In this context, the accuracy and reliability of labels play a crucial role in determining the effectiveness of the prediction model.

In summary, \name excels in handling the \problem problem compared to previous online updating approaches. Additionally, it proves effective in mitigating performance degradation over time, showcasing its robustness and reliability in real-world applications.

\subsection{RQ5: Robustness}

\begin{figure}
    \label{}
    \centering
    \includegraphics[width=0.98\linewidth]{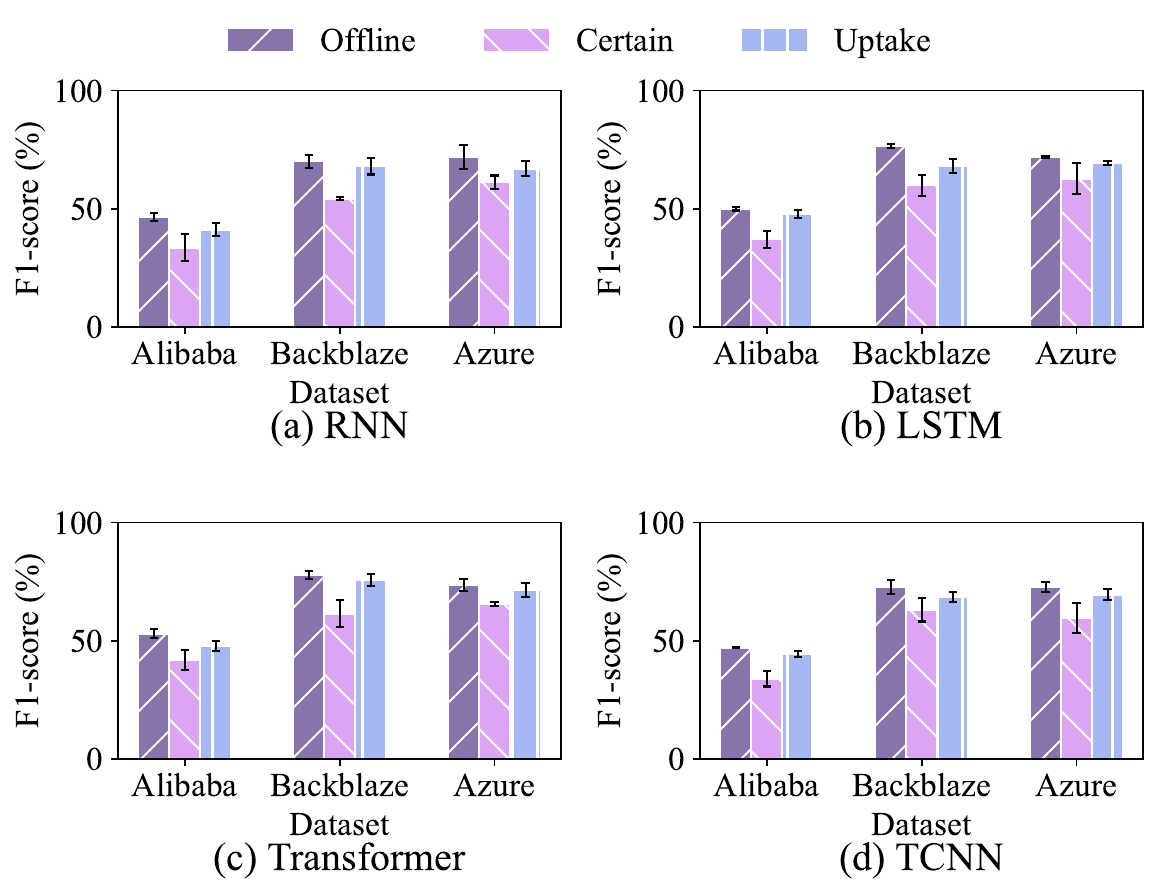}
    \caption{The average F1-score comparison of \name and its two compared approaches on three datasets: Alibaba, Backblaze, and \cloud. Each sub-figure corresponds to a specific type of base model. The error bar in the figure represents the upper and lower bounds of the results, indicating the robustness of \name across different base models.}
    \label{fig:robustness}
\end{figure}

To further empirically evaluate the performance of \name, we counted the results of \name for different base models (\ie the model in \autoref{sec:implement}) on different cloud systems (\ie Alibaba, Backblaze and \cloud). 

The provided Figure \ref{fig:robustness} displays the average F1-score across all time phases for different model update strategies on diverse base models and datasets. It is essential to clarify that "\textit{Offline}" denotes the theoretical optimal performance of the current system at a given time, assuming ideal conditions. However, achieving this theoretical best performance in practical applications is hindered by the \problem problem.

Figure \ref{fig:robustness} clearly demonstrates \name's superior performance over other model update strategies across all scenarios and models. Notably, it closely approaches the theoretical best performance (\textit{Offline}). \name exhibits greater stability over time, as indicated by the smaller fluctuations in performance metrics, highlighting its robustness and reliability compared to other model updating strategies.

\begin{table}
\centering
\caption{Efficiency comparison of \name and its two compared approaches on Alibaba, Backblaze and \cloud dataset. Each value indicates the average time (s) taken of the training epoch in the model updating process.}
\label{tab:efficiency}
\resizebox{0.7\linewidth}{!}{%
\begin{tabular}{crrr} 
\toprule
\multicolumn{1}{l}{} & \multicolumn{1}{c}{Offline} & \multicolumn{1}{c}{Certain} & \multicolumn{1}{c}{\name} \\ 
\midrule
Alibaba & 5.38 & 12.58 & \textbf{4.31} \\
Backblaze & 4.38 & 12.75 & \textbf{4.12} \\
\cloud & 11.54 & 27.90 & \textbf{10.95} \\
\bottomrule
\end{tabular}
}
\end{table}

\subsection{RQ6: Efficiency}

Besides predicting performance, efficiency is also a critical metric for online model updating. Consequently, we compare the running time of \name with other approaches proposed in \autoref{sec:experiment-compared}. Since the difference between different approaches only exists in the training step, we compared the average time cost of an epoch with the same hyper-parameters on each time phase.

From the results in \autoref{tab:efficiency}, these approaches perform almost similarly in efficiency except \textit{Certain}, and \name take the least time to train during an epoch with $4.31$s in Alibaba, $4.12$s in Backblaze, and $10.95$s in \cloud. The \textit{Certain} approach is slowest because \textit{Certain} accumulates instances with certain labels in all former phases, which leads to a larger training set. The larger the dataset it uses, the longer time it takes to train.

\begin{figure}
    \centering
    \subfigure[Alibaba]{
        \includegraphics[width=0.6\linewidth]{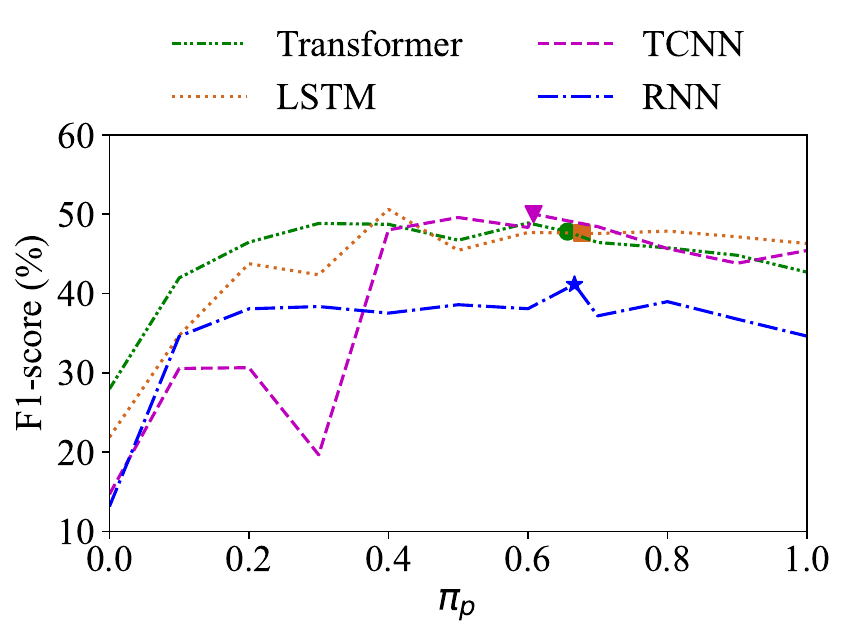}
    } 
    \subfigure[Backblaze]{
        \includegraphics[width=0.6\linewidth]{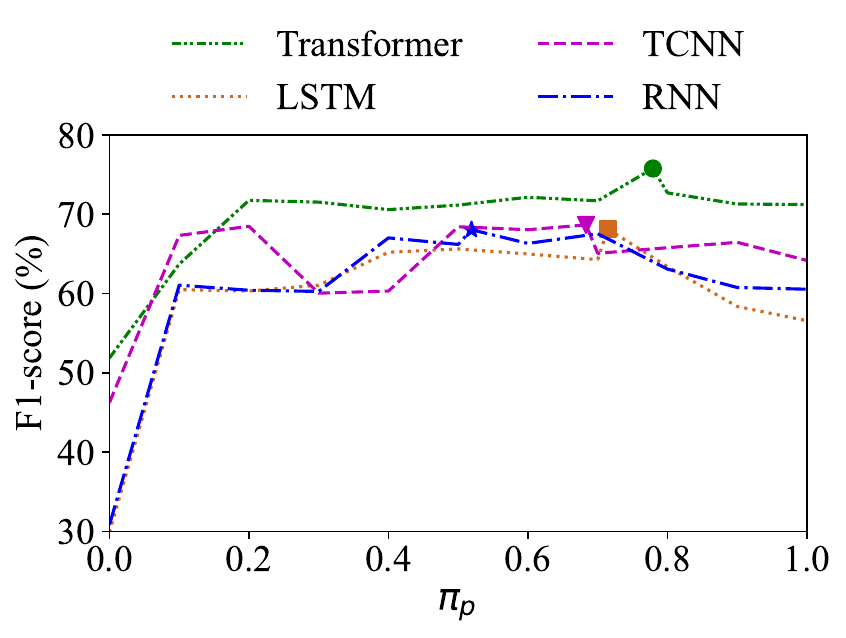}
    }
    \caption{F1-score of \name under different $\pi_{\rm p}$ on two public datasets. The dots are the parameters chosen by our solution.}
    \label{fig:configuration}
\end{figure}

\subsection{RQ7: Parameter Sensitivity}
\label{sec:experiment-parameter}

We investigate the impact of the only parameter $\pi_{\rm p}$ used in \name. 
It reflects the ratio of positive instances in all uncertain positive instances.
Our algorithm suggests using the Precision value on the T1 stage to set $\pi_{\rm p}$, which is an estimation using historical information.

In this experiment, we grid search the value of $\pi_{\rm p}$ from 0 to 1 on a 0.1 base step.
\autoref{fig:configuration} shows the effectiveness of different $\pi_{\rm p}$ in terms of F1-score. 
In this figure, each line is the average F1-score of the prediction model (RNN, LSTM, Transformer, and TCNN) on the Alibaba and Backblaze datasets, respectively. 
The dots represent the $\pi_{\rm p}$ chosen by our solution. 
Although there exists a small perturbation under different $\pi_{\rm p}$, the overall performance is stable, and our selected $\pi_{\rm p}$ always leads to almost the best performance, which indicates \name is robust in practice without the need to tune parameters carefully.

\section{Discussions}
\label{sec:discussions}

\subsection{Threats to validity}

\point{Internal threats}
Our implementation choices are a potential internal threat. To address this, we utilize established implementations for our deep learning models, detailed in Section \ref{sec:implement}. Moreover, our implementation is openly accessible, ensuring transparency and enabling future research replication.

\point{External threats}
External threats stem from model and data selection, as well as comparisons with other methods. We mitigate selection bias by incorporating diverse datasets from three distinct cloud systems, covering both node and disk-level data. These datasets align with industry standards, enhancing the representativeness of our results. Additionally, this study pioneers investigations into the \problem problem, selecting standard model updating strategies and a theoretical optimal strategy, as explained in \autoref{sec:experiment-compared}. We remain open to exploring new cloud component data and model updating techniques in future research to enhance our study's depth.

\point{Construct threats}
Construct threats concern our chosen metrics and parameters. We utilize widely-accepted metrics such as precision, recall, and F1-score for effectiveness evaluation, and training time for efficiency measurement. Future evaluations will incorporate additional metrics for a more comprehensive assessment. Parameters in \name are defined based on established rules (\autoref{sec:parameters}), with detailed discussions about the primary parameter, $\pi_{\rm p}$, in \autoref{sec:experiment-parameter}, ensuring a well-informed evaluation framework.

\subsection{Deployment}

Our \name framework is deployed on \cloud, a platform with millions of nodes serving a wide customer base. The process consists of three core phases: data preparation, model retraining, and model deployment.

\point{Data preparation}
During this phase, collected data is cleaned and engineered to ensure completeness and quality. These steps are essential for enhancing the reliability and effectiveness of subsequent model retraining.

\point{Model retraining}
In this stage, the failure prediction model undergoes retraining. To address challenges related to the \problem problem, \name is integrated into the retraining process, mitigating issues associated with the \problem problem.

\point{Model deployment}
The retrained model is deployed on \textit{AzureML}, a platform designed for seamless management and deployment of online models. Once deployed, the model actively performs online cloud failure prediction, providing insights into potential failure events within the cloud system.

To gauge \name's effectiveness in terms of business impact, we conduct A/B testing, measuring reduced mitigation action times and improved service availability. Compared to the online retraining strategy discussed in Section \ref{sec:empirical}, \name significantly reduced required mitigation actions and enhanced service availability. These results demonstrate \name's substantial benefits for online failure prediction models, highlighting its effectiveness and positive business impact.

\section{Related work}
\label{sec:related}

\point{Failure prediction}
In recent years, various approaches for predicting cloud failure have appeared, including hard drive disk failure and node failure prediction. As a binary classification problem, machine learning and deep learning mechanisms are widely used for failure prediction. Machine learning approaches for cloud failure prediction, such as support vector machine \cite{ZhaEtAl18} and tree models \cite{LiEtAl14,BotEtAl16,Huang17,XuEtAl18}, use several monitoring metrics collected in a time window to predict whether the cloud component will fail soon. However, these machine learning approaches struggle to handle the complex temporal information of cloud systems \cite{SunEtAl19}. Deep learning approaches such as RNN \cite{XuEtAl16}, LSTM \cite{lin2018predicting} and TCNN \cite{SunEtAl19} can better capture the temporal correlation of the complex monitoring metrics than classical machine learning approaches. In recent years, Transformers have outperformed conventional deep-learning approaches. The state-of-the-art performance is achieved by NTAM \cite{LuoNtam21}, which incorporates both temporal and spatial information into failure prediction. Our research is orthogonal to previous failure prediction approaches since we aim to solve the model updating issue and boost the overall performance of failure prediction.

\point{Uncertain Labels} Positive-Unlabeled Learning (PULearning) also solves uncertain/unlabeled problems for classification tasks \cite{elkan2008learning, nguyen2011positive,li2009positive, liu2002partially,li2003learning, lee2003learning,liu2003building}. 
PULearning defines positive samples and unlabeled ones for the \textit{training phase}. 
For example, an unbiased risk estimator \cite{du2014analysis} is proposed to solve this problem. 
Different from PULearning, the \problem is identified in the \textit{model updating phase}, where the samples for retraining may have three classes: positive samples, negative samples, and uncertain positive samples.

\section{Conclusion}
\label{sec:conclusion}

Cloud failure prediction is an important task to ensure the reliability of cloud systems. 
According to our real-world practice of deploying cloud failure prediction models, we identify a problem, Uncertain Positive Learning (UPLearning), during the model updating procedure. 
This problem is of great importance to tackling since it downgrades the prediction performance significantly in the real-world scenario. 
In this paper, we propose a novel model updating approach, Uncertain Positive Learning risk estimator, dubbed \name, to improve the performance of any prediction model, such as RNN, LSTM, Transformer, and TCNN.
Our experiments on both public and real-world cloud datasets demonstrate that \name can robustly achieve much better performance than the baseline approaches. 
More encouragingly, \name has been successfully applied to our cloud platforms and obtained benefits in real practice.

\clearpage

\balance

\bibliographystyle{ACM-Reference-Format}
\bibliography{references}

\end{document}